\documentclass[12pt,preprint]{aastex}

\usepackage{graphics,graphicx}

\received{2005 April 27}
\begin{document}

\title{CHANDRA Observations of the X-ray Halo around the Crab Nebula}

\author{F. D. Seward, P. Gorenstein}
\affil{Smithsonian Astrophysical Observatory, 60 Garden St., 
Cambridge MA 02138}

\author{R. K. Smith}
\affil{Goddard Spaceflight Center, Greenbelt, MD 20771}
\affil{The Johns Hopkins University, 3701 San Martin Drive, Baltimore
  MD 21218}
\begin{abstract}
Two Chandra observations have been used to search for thermal X-ray
emission from within and around the Crab Nebula.  Dead-time was
minimized by excluding the brightest part of the Nebula from the field
of view.  A dust-scattered halo comprising 5\% of the strength of the
Crab is clearly detected with surface brightness  
measured out to a radial distance of $18^{\prime}$.  Coverage is 100\%
at $4^{\prime}$, 50\% at $12^{\prime}$, and 25\% at 
$18^{\prime}$.  The observed 
halo is compared with predictions based on 3 different
interstellar grain models and one can be adjusted to fit the
obsservation.   This dust halo and mirror scattering
form a high background region which has been searched for emission 
from shock-heated material in an outer shell.  We find no 
evidence for such emission.  We can set upper limits a factor of 10-1000
less than the surface brightness observed from outer shells around 
similar remnants.  The upper limit for X-ray luminosity of an outer
shell is $\approx 10^{34}$ erg s$^{-1}$.
Although it is possible to reconcile our observation with an $8-13 M_{\odot}$
progenitor, we argue that this is unlikely.

\end{abstract}

\section{Introduction}

After 30 years of X-ray observations, the Crab Nebula remains unique
or, more accurately, peculiar when compared with other supernova
remnants.  The central Crab Pulsar accounts for $\sim5\%$ of the 1-10
keV X-ray emission.  The bulk of the emission comes from the 
surrounding pulsar-wind
nebula (PWN or synchrotron nebula) which is $\sim 2^{\prime}$ in
diameter (Bowyer et al. 1964, Palmieri et al. 1975, 
Harnden \& Seward 1984, Hester et al. 1995) and has rich, 
time-variable interior structure (Weisskopf et al. 2000, 
Hester et al. 2002).  The PWN is
surrounded by a $5^{\prime} \times 7^{\prime}$ optical nebula
comprising an array of
 He-rich filaments moving outwards with
velocities of 1000-1500 km s$^{-1}$ (Trimble, 1968, Lawrence et al, 1995).  
The mass contained in these filaments has been estimated as $1 - 5
M_\odot$ (Trimble and Woltjer 1971, Fesen et al 1997)
The kinetic
energy of this material is $2-10 \times 10^{49}$ ergs, less than
the $10^{51}$ ergs typical of other galactic and
Magellanic-Cloud remnants.   SNR 0540-69.3, in the LMC,
has a similar luminous central pulsar and PWN but, in addition, an
outer shell with L$_X \approx 8 \times 10^{35}$ ergs s$^{-1}$ and
containing $30-40 M_\odot$ (Seward \& Harnden 1994, Hwang et al. 2001).
 This emission is largely from shock-heated material
energized as the SN ejecta push through 
circumstellar gas.  This shell, which is irregular, if placed at the
distance of the Crab (2 kpc), would be $8^{\prime}-12^{\prime}$ from
the central pulsar.

Searches for emission beyond the optical filaments of the Crab have 
not yet found a convincing outer shell.  During a lunar occultation 
in 1972 a rocket flight detected soft X-ray
emission coming from outside the PWN area (Toor et al 1976).  This was
attributed to thermal emission but later shown to probably be a
dust-scattering halo.   Both Einstein and ROSAT observations
detected a faint X-ray halo extending out to $30^{\prime}$ from the
pulsar and concluded that  $\sim 10\%$ of the X-rays are in this halo
(Mauche \& Gorenstein 1989, Predehl \& Schmitt 1995).  
Because of the exceptional quality of the Chandra mirror, we thought it
worthwhile to again search for outer-shell emission.

At other wavelengths, the Crab outer shell is also elusive.
Searches by Murdin and Clark (1981) and by Murdin (1994) 
detected surrounding H$\beta$ 
emission which was thought to be the stellar wind of
the progenitor.  Fesen et al. (1997), however, showed that this emission was
widely distributed and probably not associated with the Crab.

Sankrit and Hester (1997) give 
evidence for a shock at the optical boundary of the Crab due to the 
pressure of the PWN pushing into freely-expanding ejecta located
 outside of the optical nebula.  Although the depencence of
density on radius is unknown, they estimate that several $M_{\odot}$ 
of ejecta are possible.

Sollerman et al (2000) have detected absorption in high-velocity C {\small IV}
$\lambda 1550$ and have interpreted this as absorption in fast
circumstellar material.   Parameters depend on  falloff of density
with radius and the fraction of C in the C {\small IV} state.  A shell with 4
$M_{\odot}$ and KE of $10^{51}$ ergs is possible with lower limits of
0.6 $M_{\odot}$ and $8 \times 10^{49}$ ergs using the best-fit model
with density falling off as $R^{-3}$.  In Section 6, we will consider 
this putative envelope further.  

In the radio band,
Frail et al (1995) specifically searched for an SNR shell and
found no emission out to a radius of $\approx 1^{\circ}$.
The upper limit for 333 MHz emission from any shell was 1\% of
that observed from the shell around SN 1006 -- about the same age as
the Crab and with a well-defined shell of $15^{\prime}$ radius 
($11^{\prime}$ if at 2 kpc).  A later H$\alpha$ (1410 MHz)
radio map shows a 3$^{\circ}$ diameter bubble
around the Crab (Wallace et al., 2000).  They estimate the undisturbed
ISM density as 1.6-3.5 cm$^{-3}$.

Fesen et al (1987) summarize optical studies of the Crab's
environment and review reasons for believing that there should be more
material than just the well-studied optical filaments and the
pulsar.  Current ideas of stellar evolution and collapse require that
the ZAMS precursor star have $8-13 ~M_{\odot}$.  Fesen et al estimate
the amount of material in the optical filaments to be $4.6 \pm 1.8
~M_{\odot}$. Adding a $1.4 M_{\odot}$ neutron star leaves $2-5 ~M_{\odot}$
expected to be shed in presupernova wind and high-velocity ejected material.
The interaction of ejecta with circumstellar material will produce a
shell of shock-heated gas which is readily detectable in 
X-rays from most other remnants.  The expected Crab configuration is in a
shell containing $2-4 ~M_{\odot}$ $8^{\prime}-10^{\prime}$ from the center
of the Crab (Chevalier, 1977,1985).  The present
paper describes a search for X-rays from this outer shell.  Because of the 
bright central region, scattering from the Chandra mirror, and the
bright dust halo, the search is difficult.

\section{Chandra Observations}

Hester et al. (2002) observed the Crab Nebula 8 times  
from November 2000 to April 2001.  They used an 
ACIS-S subarray to minimize pileup in the detector.  The field of view
was $2.4^{\prime}\times8^{\prime}$, enough to include the brighter 
parts of the PWN and to study time-variation of this
structure.  Because of limited telemetry response, the effective
exposure of these 25 ks observations was only 4 ks, a factor of
6  dead
time.  To avoid this problem, we excluded the bright central region
from our observations.  Our first observation was a 20 ks exposure
using the 4 ACIS-I chips and pointed $10^{\prime}$ N of the pulsar.  The
X-ray nebula was not in the field of view.  The second observation was
a 40 ks exposure using 3 ACIS-S  and 2 ACIS-I chips with the 
X-ray nebula centered on the S3 chip, but with the center region of
the chip excluded from the telemetry.  Thus, with dead time only a
few percent, 20 and 40 ks exposures were
obtained of the halo and of the faint outer part of the PWN. Table 1
gives detail for these 2 observations and includes one of the shorter
subarray observations.

Figure 1 shows the sum of these 3 observations in the energy 
range 0.4--2.1 keV.  To show the inner nebula,
one of the 4 ks observations of the bright Crab has been normalized
and added to fill the hole left by telemetry exclusion.  Cosmic-ray
background has been subtracted and time variation of chip sensitivity has
been included.  The ACIS charge-transfer streak
has been subtracted from chip S3, on which the Crab is imaged, and from
the 2 I chips north of and overlapping the field of S3.  In order to
show detail at the center, some chips are only partially shown in this
figure.  Since calibration of some
chips is more extensive than for others, chip IDs are listed.

There is appreciable structure at the outer boundary of the PWN.
The faintest features visible are $2.5^{\prime}$
 from the center of the nebula and
these have surface brightness a factor of 200 less than that where the PWN
is brightest.  The halo data extend from this radius, which is inside
the optical nebula, to a radial distance of $18^{\prime}$.  In this
span, the halo brightness decreases a factor of 100.  Coverage of the
halo is 100\% at radial distances from $2.5^{\prime}$ to $4^{\prime}$, 
is greater
than 60\% out to $10^{\prime}$, and falls to 25\% at $18^{\prime}$. 

Figure 2 shows measured surface brightness extending from the center
of the Crab to the outermost chip boundaries.  Data from the central,
northern, and western chips indicate a halo with intensity independent
of azimuth.  The 2 southern chips, S2, and S1, show greater surface 
brightness.
This is, at least partially, a calibration problem.  Excluding these 2
chips, the halo is symmetric about the point: RA = 05 34 31.3, 
Dec = 22 01 03,
located $14^{\prime\prime}$ northwest of the pulsar and within the bright
X-ray torus. 

Figure 3 shows the Chandra-measured surface brightness compared to that
measured by ROSAT (Predehl and Schmitt 1995).  The energy range of
both observations is $\approx$ 0.4-2.1 keV but the ROSAT sensitivity
from 1.5-2.1 keV is considerably less than that of Chandra.  To obtain
the strength of the Chandra dust halo, the Chandra mirror scattering
(dashed line) must be subtracted from the observed brightness (solid
line).  Note that the Chandra measurement, even before this correction,
falls below the ROSAT observation.  The mirror scattering was taken
from an observation of Her X-1 combined with ground calibration as
summarized by Gaetz (2004).
The strength of the Chandra dust halo integrated from $2.5^{\prime}$ 
to $18^{\prime}$ and interpolated from $0^{\prime}$ to $2.5^{\prime}$, 
is 0.047 that of the Crab Nebula.  (The interpolation from 
$0^{\prime}$ to $2.5^{\prime}$ accounts for .010 of this.)  The
ROSAT-measured scattered fraction from $0^{\prime}$ to $30^{\prime}$
is 0.080 (Predehl \& Schmitt 1995).  An extension of our
Figure 3 curve to $30^{\prime}$ would increase the
Chandra-measured fraction to $0.048 \pm 0.008$. Uncertainties comprise
measure of surface brightness, measure of total crab count rate,
extrapolation to small radii, background subtraction, and an assumed
20\% error in the mirror scattering. 

We note that we have used data
taken N and E of the Crab and we have assumed that this is valid
for all azimuths.  We have not included data from the 2
southern ACIS chips because the discontinuity at the chip boundary
indicates a normalization problem.  If we assume that the higher 
surface brightness indicated by these chips (S1,S2)
is real, and that this higher brightness applies to a sector 
extending $90^{\circ}$ in azimuth,
the Chandra-observed scattered fraction to $30^{\prime}$ would
increase to 0.051, which is within our margin of error.  Since the
ROSAT and Chandra detectors have different spectral sensitivities, even
though the energy range covered here is about the same as that of
ROSAT, the fraction of counts in the halo is not expected to be the
same.  Using the known spectrum of the Crab and the halo spectra
given here at the end of Section 4, we expect the relative strength of
the dust halo measured with Chandra to be 82\% of that measured with
ROSAT (because the ROSAT detector is relatively more sensitive at 
low energies).  We observe a halo strength $60\% \pm 10\%$ that of
ROSAT so conclude that the ROSAT result is too high.

Figure 4 was made to illustrate fluctuations in halo surface
brightness.  The 0.2-2.1 keV data shown in Figure 1 was first smoothed
to make map $M$.  Then a function $F(r)$, with about the same radial 
dependence of surface brightness was subtracted. 
$F(r) = constant [ 1 + (r/240)^2]^{1.05}$, where $r$ is distance from
the scattering center in ACIS pixels.  The figure shows the quantity
$(M-F)/F$ and one can see regions N and S of the Crab which are $\sim
10\%$ brighter than average.  Note that since the average decrease
of brightness with radial distance has been removed from Figure 4, any
extended above-average component also is decreasing with radial
distance, contrary to appearance in Figure 4.  We interpret the
significant features in Figure 4 as possible
structure in the dust distribution and/or variations in column density
of absorbing gas in the line-of-sight.
The $1^{\prime}-2^{\prime}$ feature $5^{\prime}$ SSE of the Crab
center is discussed further in section 4.  Note that any gradual
radial variation in brightness implied by Figure 4 may be an artifact
due to the form assumed for the subtracted function, F(r).  Apparent
azimuthal variation should be real.  Because we are seeing variations
of a few percent, chip-to-chip calibration uncertainties show.  A
variable contamination layer on the instrument window is also
a cause for concern.  This layer, however,  is thicker at the edges 
of the window and, if present, should produce a recognizable effect.  
This is not seen.

The halo spectrum contains no strong sharp features which might
indicate thermal emission from a shock.  Reasonable fits
are obtained using the sum of power law and thermal bremsstrahlung
(used as an arbitrary continuum) components.  
The signal is comprised of dust-scattered halo, mirror
scattering, and background, which is negligible except for
high energies at large angles.  
 
\section{Upper limits to outer shell}

To be detectable, X-rays from any shock-heated material must be
visible over the dust-scattered halo.  Since diffuse uniform emission is
more difficult to detect than bright knots, we consider 
a hypothetical diffuse shell which represents the most massive 
allowed shell possible.  The
limiting surface brightness is taken as 0.1 of the observed dust halo.

Upper limits depend on radius, $R$, of the assumed shell and were
calculated assuming a spherical shell of thickness $0.15R$ centered on
the pulsar and filled with material of uniform density, $n$ 
cm$^{-3}$.  Using the dashed curve of Figure 5 (0.1 of the observed
halo) as the surface brightness of the unseen shell, limits on several 
quantities are calculated and
shown in Figure 6.  The upper limit to $n$ is 4 just inside the
optical nebula and drops to 0.15 at $R=18^{\prime}$.  If the
surrounding ISM is uniform, since $n$ is swept-up material, the ISM
density, $n_0$ would be 0.4 of these values.  The
limit on the X-ray luminosity, $L_x$, of any shell is $\approx
10^{34}$ ergs s$^{-1}$ and almost independent of $R$.  Uncertainty of
the gas temperature leads to an uncertainty of $\pm 25\%$ in $n$ and
$\pm 40\%$ in $L_x$.  The calculation of $n$ and $L_x$ is
straightforward.  A model is necessary to derive paremeters of the
explosion.  It is customary to estimate the energy of the shock,
$E_{\circ}$, using a simple blast-wave model (Cox, 1972).  For a
uniform ISM, $E_{\circ}/n_0 = 1.6 \times 10^{-6} R^5 t^{-2}$, where 
the units of $E_{\circ}$ are $10^{51}$ ergs, $R$ is in
parsecs and the age, $t$ is in $10^4$ years or, in this case, $t =
.095$.  Upper limits for $E_{\circ}$ are shown in Figure 6.  

The crosses in Figure 5 show measured surface brightness of selected
``bright spots''.  These illustrate that the limit of bright knot 
detectability is about 0.1 the brightness
of the dust halo.  All are consistent with statistical fluctuations 
except for the point at $R = 2.6^{\prime}$ which is a small cloud of
emission within the N boundary of the optical nebula. 
 At $R = 12^{\prime}$ the bright lumps
represents a knot size of $\sim 1$ pc and a lump luminosity of $3 \times
10^{28}$ ergs.  Assuming we would notice 10 such lumps in a
$30^{\circ}$ arc, this would imply 300 lumps in the shell, a total 
$L_x = 10^{31}$ ergs and total mass of $2 \times 10^{-2}$ M$_{\odot}$.
As expected, these limits are far below the limits calculated for a
diffuse uniform shell.

The circles in Figure 5 show surface brightness of shells observed
by Chandra in other remnants (Seward et al 2004).  
Most remnants have an irregular outer shell which defines the boundary
and brighter patches at a lesser radius.  In this figure, we have shown
brightness and radial position for both the brightest part of the
shell and the emission observed over most of the outer boundary.
Radii have been corrected to show the size at 2 kpc distance.  
Although surface brightness does
not depend on distance, corrections have been made for differing
absorption measured in the ISM.  The remnants Kes 75 and SNR 0540-69.3
have bright central PWN very similar to that of the Crab and, in this
respect, are the most Crab-like remnants known.

We searched, without success, for thermal emission inside the optical
nebula.  There are many faint features at the edge of the PWN.  All have
soft power-law spectra and are best interpreted as part of the PWN.
The density of any unseen thermal X-ray-emitting diffuse material 
must be $< 4$ and the mass $< 0.2 M_{\odot}$.  The limits on lumpy
material are appreciably less.

\section{Dust scattering}

Although no emission from an outer shell has been recognized, there is
substantial extended emission observed due to scattering from dust in
the interstellar medium (ISM) and mirror scattering in the Chandra
HRMA.  As we will show, below $\approx$ 2.5 keV scattering by dust grains
dominates the extended emission; above $\approx$ 3 keV mirror
scattering becomes the primary contribution.  

X-ray scattering by ISM grains, first
described by Overbeck (1965), has been observed by instruments on
Einstein (Mauche \& Gorenstein, 1986), ROSAT (Predehl \& Schmitt,
1995), Chandra (Clark, 2004, Smith, Edgar, \& Shafer, 2002), and XMM 
(Vaughan et al. 2004).
Theoretical studies have been done by Mathis \& Lee (1991),
Predehl \& Klose (1996), and Smith \& Dwek (1998). 

The total scattering cross section in the Rayleigh-Gans
(RG) approximation illustrates the dependence on X-ray energy and
grain characteristics.  It is applicable when $E > 2$\,keV and is
\begin{equation}
\sigma(E, a) = 6.3\times10^{-7} \Big({{2Z}\over{M}}\Big)^2
  \Big({{\rho}\over{3 \hbox{g\,cm}^{-3}}} \Big)^2 a_{\mu\rm m}^4
  E_{\rm keV}^{-2} {\rm cm}^2
\label{eq:intRG}
\end{equation}
where $a$\ is the grain radius, $Z$\ is the mean atomic charge, $M$\
the mean atomic weight (in amu), $\rho$\ the mass density, and $E$\
the X-ray energy in keV (Mathis \& Lee 1991).  Eq.~\ref{eq:intRG}\ implies
that the overall scattering halo will tend to be brighter at lower
energies, from the $E^{-2}$\ term [note error in Mathis \& Lee (1991) 
showing this as $E^2$].  Figure~\ref{fig:TotScatFrac}\
plots the total scattering fraction between $120-1000''$, the range
observed here, assuming a column density of N$_{\rm H} =
10^{21}$\,cm$^{-2}$.  Three different dust models, Mathis, Rumpl, \& 
Nordsieck (1977) (MRN), Weingartner \& Draine (2001)
(WD01; using $R_V = 3.1$\ and $b_C = 6\times10^{-5}$), and
Zubko, Dwek \& Arendt (2004) (ZDA04; using the BARE-GR-B parameters) 
are shown using
both the exact Mie solution for scattering from a sphere and the
approximate RG solution.  In all cases the RG approximation clearly
begins to break down below 1.5 keV, although the scattering is generally
larger at lower energies.  The ZDA04 model, which has relatively fewer
large grains than the MRN and WD1 models, gives the best fits of the
three to our data (see Figure 8).

The analysis to be described used only data from the 4 I chips of the
14 April 2002 observation (obsid 2798).  There was a
charge-transfer  streak in chip I0 due to part of the Crab PWN at
the edge of the chip.  The charge transfer streak was 
therefore subtracted from the 2 chips closest to the Crab.  For
each energy interval, the counts were projected along the transfer
axis and summed. 0.013 of this sum was then subtracted from each
element of the image.

At almost any energy, extracting an X-ray scattering halo from the
observations first requires that the Chandra PSF be subtracted.  As
described by Smith et al. (2002), ray-trace models of the Chandra PSF
(such as ChaRT) significantly underestimate the scattering at angles
beyond $1^{\prime}$.  Therefore we followed
Smith et al. (2002) and used an on-axis Her X-1 observation (obsid 3662)
as our PSF calibrator.  This has the obvious limitation that this
observation was done on-axis, while our Crab observation was done with
the Crab $\sim 10'$\ off-axis.  We believe that this is reasonable
because at 4 keV, where dust scattering is minimal, the observed Crab
profile matches the Her X-1 profile.  We note, however, that while
this match is suggestive it does not guarantee that there are no
differences in the PSF at lower energies.

Unlike most halo studies, the Crab nebula is not a point source but
rather an extended nebula $\sim 1'$\ in radius.  We calculated the
radial profile assuming it was centered at 05:34:31.3, 22:01:03
(J2000), which is both roughly central and near the peak of the
nebular emission.  This is not the location of the Crab pulsar,
however, which itself emits only 5\% of the X-ray emission.  The
effect of source extent is relatively minor
except at scattering angles comparable to the size of the source.
With the assumption that the source is circular with uniform surface
brightness, the effect can easily be calculated by integrating the point-source
scattering intensity over the surface:
\begin{equation}
I(\theta, \phi) = 2 \int_{\theta-\phi}^{\theta+\phi} d\psi \psi
\arccos((\theta^2 - \phi^2 + \psi^2)/(2 \theta \psi)) I(\psi)
\label{eq:finitesize}
\end{equation}
where $\phi$\ is the source radius on the sky and $I(\theta)$\ is the
scattered halo at angle $\theta$.  This equation holds for $\theta >
\phi$; in most cases, when $\theta < \phi$\ the source brightness itself 
will swamp the scattered halo.  

We extracted the radial profile of the Crab Nebula in energy slices
between 0.5-4 keV.  Between 0.5-1.0 keV, we used an energy width of
0.1 keV (approximately equivalent to the energy resolution of the ACIS
CCDs), and between 1.0-4.0 keV we used a width of 0.2 keV.  
We modeled the Crab
as a uniform circle of radius $1'$, and fit it using various dust
models using Eq.~\ref{eq:finitesize}\ and either the Mie solution (for
energies below 1.5 keV) or the RG approximation (above 1.5 keV).
Sample results at 1 and 2 keV, assuming the dust has an MRN-type size
distribution and is smoothly distributed between the Crab and the Sun
are shown in Figure~\ref{fig:SampleHalo}.

As Figure~\ref{fig:SampleHalo}\ shows, by 2 keV the observed radial profile is
strongly influenced by the power-law shape of the PSF; at 1 keV, the
shape of the observed profile shows dust scattering is dominant.  The
1 keV X-ray surface brightness is poorly fit by the MRN model.  Changing the
assumed dust model to a WD01 or ZDA04 model does not significantly
improve the fits.

If the dust is assumed to be smoothly distributed along the line of 
sight, the choice of a dust grain model leaves only the total dust 
column density as a free parameter; this can easily be converted to a 
gas column density using the dust model parameters.  In Figure 9 we 
show the best-fit hydrogen column density for the three different dust 
grain models as a function of energy. Since the energy dependence of 
the halo emission has already been taken into account in the model 
fits, any variation with energy indicates the model does not completely 
describe the data. Figure 9 shows that the best-fit column density from 
the halo data is significantly lower than the best-fit column density 
derived from fitting the X-ray spectrum, $N_H \approx
3.5\times10^{21}$\ atoms cm$^{-2}$.  This result disagrees with that of 
Predehl \& Schmitt  (1995) but is consistent with our observation of 
less halo emission than they saw with ROSAT.

Regarding the variations seen in Figure 9, an examination of the 
individual halo fits showed that this simple "smoothly distributed 
dust" model fit best at energies between 1.5-2.5 keV.  At higher 
energies, we believe that errors in the mirror scattering model 
dominate the fits.  At lower energies, it seems likely that the 
one-component model is too simple, as described below.  We also note 
that the error bars in Figure 9 are purely statistical, and do not 
include the known but difficult-to-estimate systematic errors such as
the energy dependence of the Chandra mirror point-spread function.

To improve the fits, we experimented with more complex models, with
two halo components: a ``smooth'' component plus a single cloud of
dust between the Sun and the Crab.  In this case, we find reasonable
fits, although the column density varies a bit with energy.  We find
that the planar dust to be very near, with a column density of
$\sim 4-5\times10^{20}$\,cm$^{-2}$, while the smooth dust has a column
density of $\sim 8-9\times10^{20}$\,cm$^{-2}$\ for MRN-type dust.  If
instead we use a ZDA04 dust model (specifically their
BARE-GR-B model), as shown in Figures 10 and 11,
we get significantly improved fits over a MRN-type
distribution.  Again, this column density is lower than normally used
for the Crab, and is affected by the dust size distribution chosen.

Interestingly, the Local Bubble (LB) radius is, on average, $\sim 100$\,pc
distant (Cox \& Reynolds, 1987).  Assuming an ``average'' IS density of 1
cm$^{-3}$\ existed before the LB was swept out implies the edge would
have a column density $\sim 3\times10^{20}$\,cm$^{-2}$.  Observations
of the LB edge by Lallement et al. (2003) show that the edge in the
direction of the Crab is at $\sim 200$\,pc, with a column density
greater than $10^{20}$\,cm$^{-2}$.  

Although plausible, we cannot conclude that this excess at large
angles is due to the LB edge.  It could also be caused by additional
small dust particles that are not in the model, or even due to a
missing mirror scattering term.  In addition, at these large angles
the data is from the outer two CCDs.  Therefore there is no blurring
correction from the bright edge of the nebula, although calibration
differences between the various chips could contribute to the excess
as well.

In sum, our primary results concerning dust are:
\begin{itemize}
\item The ZDA04 model seems to best fit the radial dependence of
  surface brightness.
\item There appears to be less dust along the line of sight to the
Crab than would be predicted from the best-fit N$_{\rm H}$ value for
the Crab spectrum, although
this may depend on the dust model used.
\item There is evidence for a nearby plane or cloud of dust with a moderate
  column density. 
\end{itemize}

Figure 12 shows the spectrum of the halo $6.5^{\prime}$ SSE of the
scattering center.  The mirror scattering is approximated by a broken
power law with indices 1.1 and 2.8 and a break at 4.6 keV.
All events with energies above 2.5 keV are assumed to be from the
mirror.  The dust contribution below 2.5 keV 
was approximated and characterized by
a continuum.  Of the several simple models readily available, a
bremsstrahlung spectrum gave the best fit with about the right value for
$N_H$.  No emission mechanism is implied.  The
residuals to halo spectra typically
show a multiple peaked structure between 0.8 and 2 keV.
This structure, which varies from place to place and is about 5\% of
the signal at most locations, is not understood.
Adding models with line emission does not produce reasonable fits.
Some of the structure may be an artifact of the detector.
For example, some spectra contain a line feature at 1.5 keV which
probably comes from an Al coating on the detector window.
In any case, the ``temperature'' of the bremsstrahlung continuum
characterizes the dust-scattered spectrum.  Some results at varying
distances are: $4.5^{\prime}$, 0.48 keV; $6.5^{\prime}$, 0.37 keV;
$8.5^{\prime}$, 0.32 keV; $15^{\prime}$, 0.23 keV.  As expected, 
the scattered spectrum is softer as scattering angle increases. 

\section{Nearby sources}

The Chandra mirror is well suited for the detection of point sources
embedded in diffuse emission.  There are 19 serendipitous
sources visible to the eye in the field shown in Figure 1. 
Because of smoothing, compression, and color map, only one is 
visible (barely) at the western edge of Figure 1 but shows clearly in
Figure 4. 
The closest
source to the Crab Nebula is at RA = 05 34 45.91, Dec = 22 00 11.6 (2000). 
This is $3.3^{\prime}$ from the pulsar and on the eastern boundary of
the optical nebula.  Strengths range from 1 to 12 counts ks$^{-1}$ and
none fall clearly within the projection of the optical nebula. 

\section{Discussion/Conclusions}

There is no indication in our observation of X-ray 
emission from an outer shell.  The shell predicted assuming the
expected type II SN progenitor has $\approx 4 M_{\odot}$ and is moving at
$\approx 5000$ km s$^{-1}$.  If the ``usual'' blast wave analysis of
Section 3 is
done, we conclude that this shell does not exist. At a radius of $R =
10^{\prime}$ a uniform shell containing $\approx 2 M_{\odot}$ and
indicating an explosion energy of $10^{50}$ ergs is possible but highly
unlikely.  All other remnants which have prominent outer shells are
irregular.  If the Crab outer shell were similarly clumpy, limits on
emission, would be considerably
lower than the limits used here.  Our upper limits for emission are 
already a factor of 100 - 1000 below that observed from shells around
SNR 0540-69.3 and Kes 75 which have small bright PWNe similar to the
Crab.  Even the weak plerionic remnant G21.5-0.9, with central
pulsar and surrounding PWN ($70 \times$ less
luminous than the Crab) has 2 shell-like features which, as shown in
Figure 5, are still $\sim$ 10 times brighter than our limit.  

At radii $> 10^{\prime}$, a larger mass and energy are possible and
our coverage becomes sparse.  ROSAT and Einstein observed out to
$30^{\prime}$ with 100\% coverage and found no shell-like emission:
so we know there is no bright shell just outside the Chandra field of view.
A faint shell is possible.

The freely-expanding ejecta proposed by Sankrit and Hester (1997)
and by Sollerman et al (2000) consists of
photoionized $10^4 -10^5$ K material and is too cool to be  
detected by Chandra.  Shock-heated material, however, will be present 
where this fast moving ejecta plows
into the pre-supernova environment.  This would be detectable by
Chandra if the density
of the shocked material were high enough.  The Sollerman et al shell
density varies as $R^{-3}$; our upper limit varies as $R^{-2}$.
Assuming a shock structure similar to that given by Chevalier (1982,
Figure 2), the reverse shock in the ejecta should have a density 4
$\times$ that in the unshocked material.  For the Sollerman et al 
minimum-mass model, this is above our limit at
$R < 6^{\prime}$.  The shock in the presupernova ISM, assuming a
similar density jump, would be below our limit at all $R < 18^{\prime}$ if
$n_0 < .02$.

In conclusion, with reasonable assumptions about non-uniform
distribution and density, we find no evidence for the shell expected
from an $8 - 13 M_{\odot}$ SN in the region $2^{\prime} < R <
\approx 8^{\prime}$, where the velocity of freely-expanding material ranges
from $\approx 1200$ to $\approx 4800$ km s$^{-1}$.
We cannot exclude models postulating several $M_{\odot}$ of 
ejecta with temperature $10^4-10^5$ K if the circumstellar
density is very low ($\sim .01$) and rather uniform.  We note that
quantitave comparison with these models is very uncertain

Although our X-ray upper limit is an order of magnitude lower than
past work, we cannot firmly exclude a $10^{51}$ erg explosion of a
$8 - 13 M_{odot}$ progenitor.
Certainly the range of possible
circumstances is narrowing.  Any hidden mass is almost invisible.
We note that 3C 58 (Slane et al 2004) and G054.1-0.3 
(Lu et al 2002) have central pulsars and PWN but
have weak (or absent) X-ray emitting shells.  Although both only 
$1.5 \times 10^{-3}$ as luminous as the Crab, these, together with the Crab, 
 may form a class of gravitational-collapse SNe with unusual progenitors.

This work was supported by Chandra Grants GO2-3087X and GO4-5059X.  We
thank Rob Fesen for a critical reading of the manuscript, for
several important references, and for showing enthusiasm over a
non-detection observation.

\section{References}

\def\ref#1  {\noindent \hangindent=24.0pt \hangafter=1 {#1} \par}

\ref{Bowyer, S., Byram, E., Chubb, T., \& Friedman, H., 1964, Science
  146, 912}
\ref{Brinkman, W., Aschenbach, B. \& Langmeier, A., 1985, Nature
  313, 662}
\ref{Clark, G.W. 2004, ApJ 610, 956}
\ref{Cox, D.P., \& Reynolds, R.J.\ 1987, ARA\&A, 25, 303}
\ref{Chevalier, R.A., 1977, in Supernovae, edited by D. N. Schramm
  (Reidel, Dordrecht), p53.}
\ref{Chevalier, R.A., 1982, ApJ 258, 790}
\ref{Chevalier, R.A., 1985, in The Crab Nebula and Related Supernova
  Remnants, edited by M. C. Kafatos and R. B. C. Henry (Cambridge
  University Press, Cambridge), p. 63.}
\ref{Fesen, R.A., Shull, J.M., \& Hurford, A.P., 1997, AJ 113, 354}
\ref{Frail, D.A., Kassim, N.E., Cornwell, T.J., \& Goss, W.M., 1995,
  ApJ 454, L129}
\ref{Gaetz, T.J., 2004, Chandra X-ray Center memorandum, 23 June 2004}
\ref{Harnden Jr., F.R. \& Seward, F.D., 1984, ApJ 283, 279}
\ref{Hester, J.J., et al., 1995, ApJ 448, 240}
\ref{Hester, J.J., et al., 2002, ApJ 577, L49}
\ref{Hwang, U., Petre, R., Holt, S.S., \& Szymkowiak, A.E., 2001, ApJ
  560, 742}
\ref{Lallement, R., Welsh, B.Y., Vergely, J.L., Crifo, F., \& Sfeir,
  D., 2003, A\&A 411, 447}
\ref{Lawrence, S., MacAlpine, G., Uomoto, A., Woodgate, B., Brown, L.,
  Oliversen, R., Lowenthal, J., \& Liu, C. 1995, AJ 109, 2635}
\ref{Lu, F.J., Wang, Q.D.,Aschenbach, B., Durouchoux, P., \& Song,
 L.M., 2002, ApJ 568, L49}
\ref{Mathis, J.S., Rumpl, W. \& Nordsieck, K.H. 1977, ApJ 217, 425}
\ref{Mathis, J.S. \& Lee, C.W. 1991, ApJ 376, 490 }
\ref{Mauche, C.W. \& Gorenstein, P., 1986, ApJ 302, 371}
\ref{Mauche, C.W. \& Gorenstein, P., 1989, ApJ 336, 843}
\ref{Murdin, P. \& Clark, D.H. 1981, Nature 294, 543.}
\ref{Murdin, P., 1994, MNRAS 269, 89}
\ref{Overbeck, J. W. 1965, ApJ 141, 864}
\ref{Palmieri, T.M., Seward, F.D., Toor, A., \& Van Flandern, T.C., 1975,
ApJ 202, 494}
\ref{Predehl, P. \& Schmitt, J.H.M.M., 1995, A\&A 293, 889}
\ref{Predehl, P., \& Klose, S., 1996, A\&A 306, 283}
\ref{Sankrit, R. \& Hester, J. 1997, ApJ 491, 796}
\ref{Seward, F.D. \& Harnden, F.R., 1994, ApJ 421, 581}
\ref{Seward, F., Slane, P., Smith, R., Gaetz, T., Lee, J.J., Koo,
  B.C., 2004, http://snrcat.cfa.harvard.edu}
\ref{Slane, P.O., Helfand, D.J., van der Swaluw, E. \& Murray, S.S.,
  2004, ApJ 616, 403}
\ref{Smith, R.K. \& Dwek,~E. 1998, ApJ 503, 831}
\ref{Smith, R.K., Edgar, R.J., \& Shafer, R.A., 2002, ApJ 581, 562}
\ref{Toor, A., Palmieri, T.M., \& Seward, F.D., 1976, ApJ 207, 96}
\ref{Trimble, V., 1968, AJ 73, 535}
\ref{Trimble, V. \& Woltjer, L., 1971, ApJ 163, L97}
\ref{Vaughan, S., et al. 2004, ApJL 603, L5}
\ref{Wallace, B.J., Landecker, T.L., Kalberla, P.M.W., \& Taylor,
  A.R., 1999, ApJS 124, 181}
\ref{Weingartner, J.C. \& Draine, B.T., 2001, ApJ, 548, 296}
\ref{Weisskopf, M.W. et al., 2000, ApJ 536, L81}
\ref{Zubko, V, Dwek, E. \& Arendt, R.  2004, ApJS 152, 211}
\newpage


\newpage

\begin{figure}
\center
\plotone{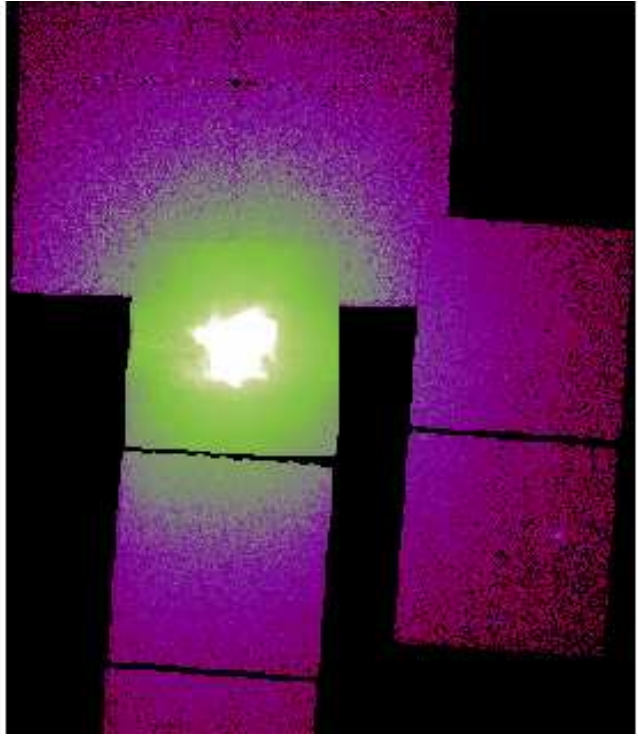}
\caption{Summed Chandra observations in the range 0.4-2.1 keV 
showing the bright nebula and faint halo.  Data have been smoothed
with a Gaussian of $9^{\prime\prime}$ FWHM.  Some ACIS chips are only
partly shown in this figure.  Reading left to right, top to bottom 
(like a book), the chips are: I3, I1, I2, I0, S3, I3, S2, I2, S1.  }
\end{figure}

\begin{figure}
\center
\plotone{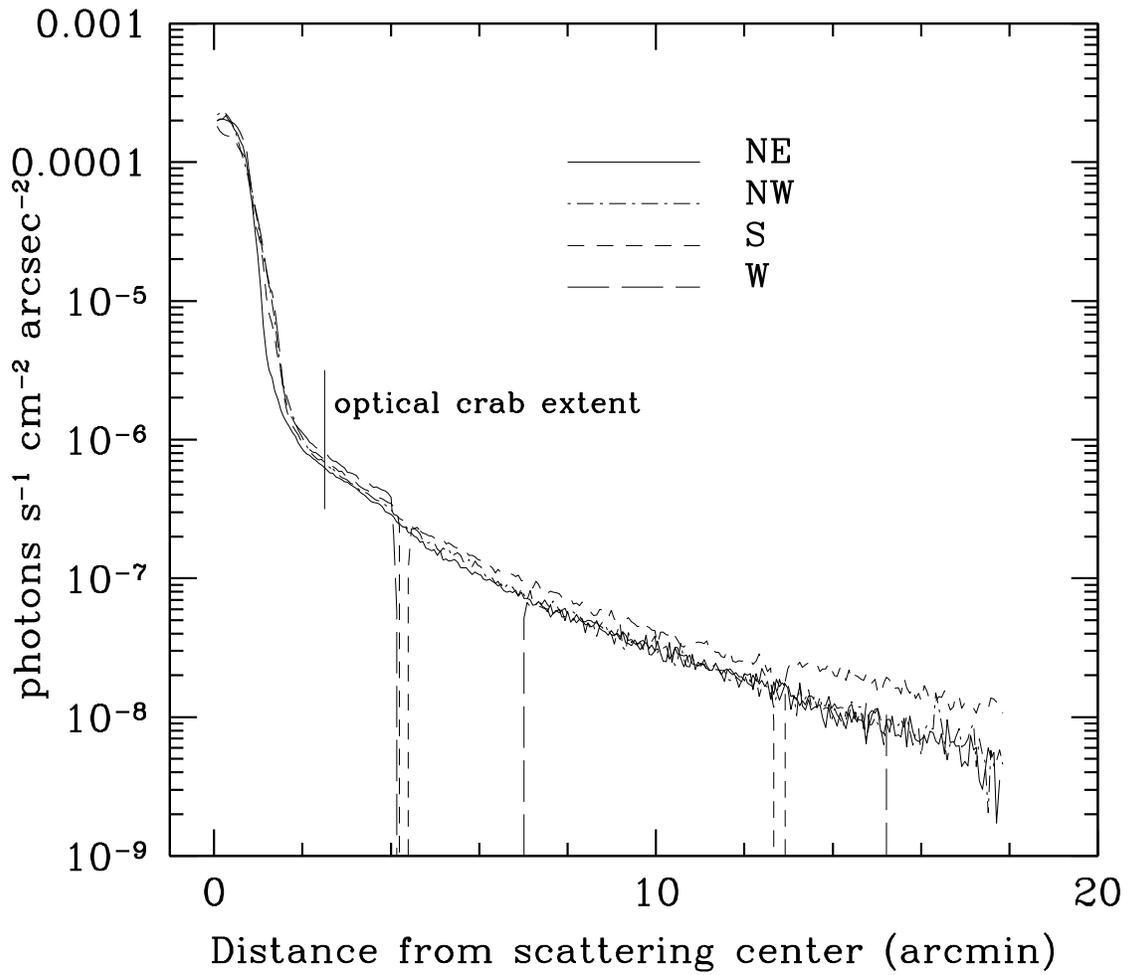}
\caption{Measured surface brightness in 4 directions.  Vertical
  lines show edges of the ACIS chips.  Data closer than
  $4^{\prime}$ are all from chip S3; beyond $4^{\prime}$, data are
  from 7 different chips}
\end{figure}

\begin{figure}
\center
\plotone{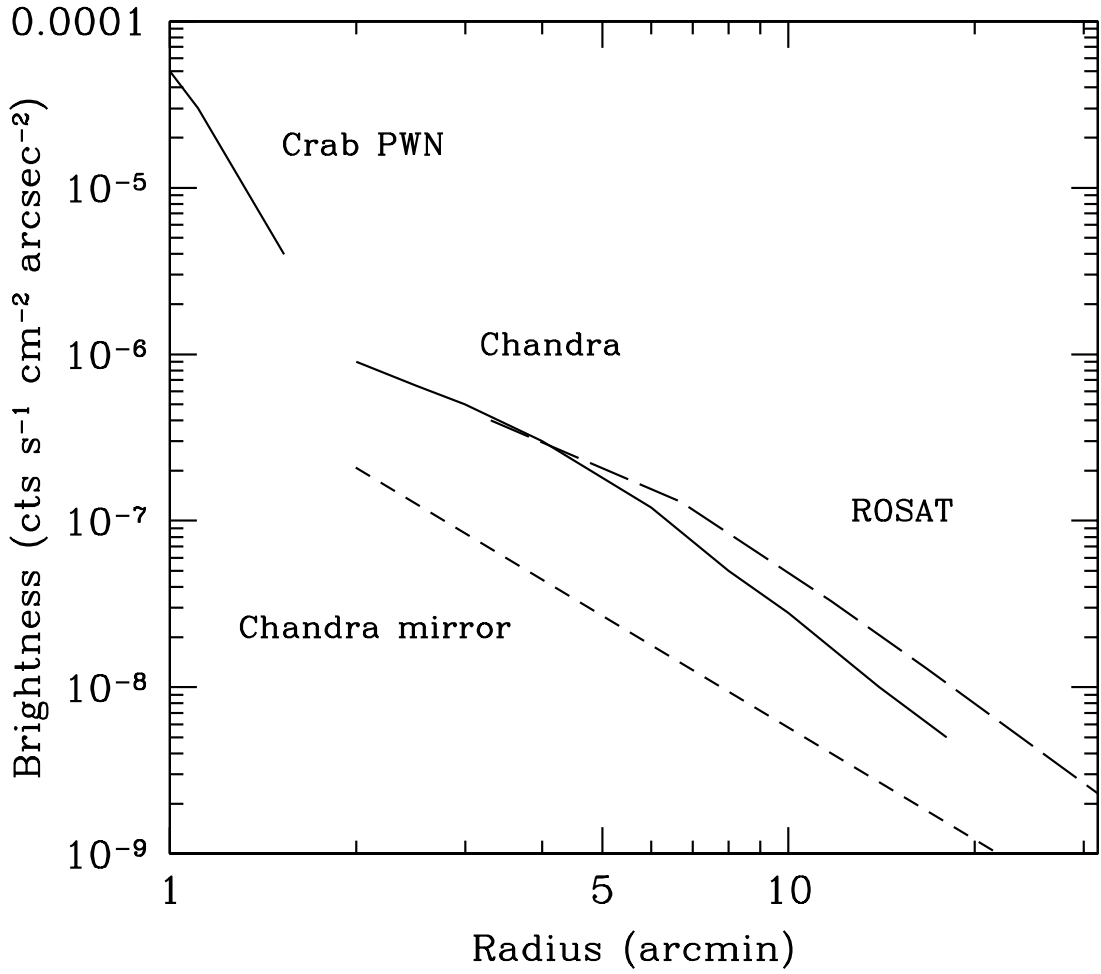}
\caption{Dust halo surface brightness measured by Chandra and 
ROSAT.  Mirror scattering has been subtracted from the ROSAT data but
not from the Chandra data.}
\end{figure}

\begin{figure}
\center
\plotone{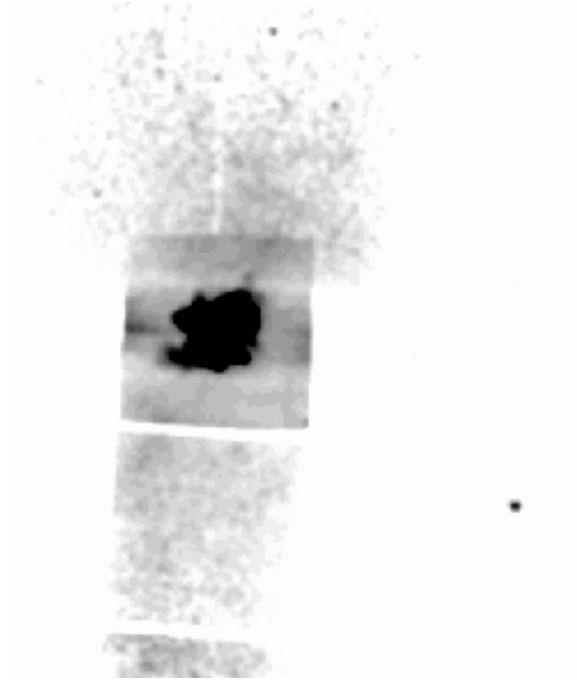}
\caption{Regions of above average surface brightness.  Generation
of this figure is described in Section 2.  Horizontal bands in the
central chip,
S3, show imperfect subtraction of the charge-transfer streak. The
brightness of the halo in the southernmost chip, S1, and the weakness 
of the halo in the 2 western chips could indicate that the relative chip 
normalization is not quite correct.}
\end{figure}

\begin{figure}
\center
\plotone{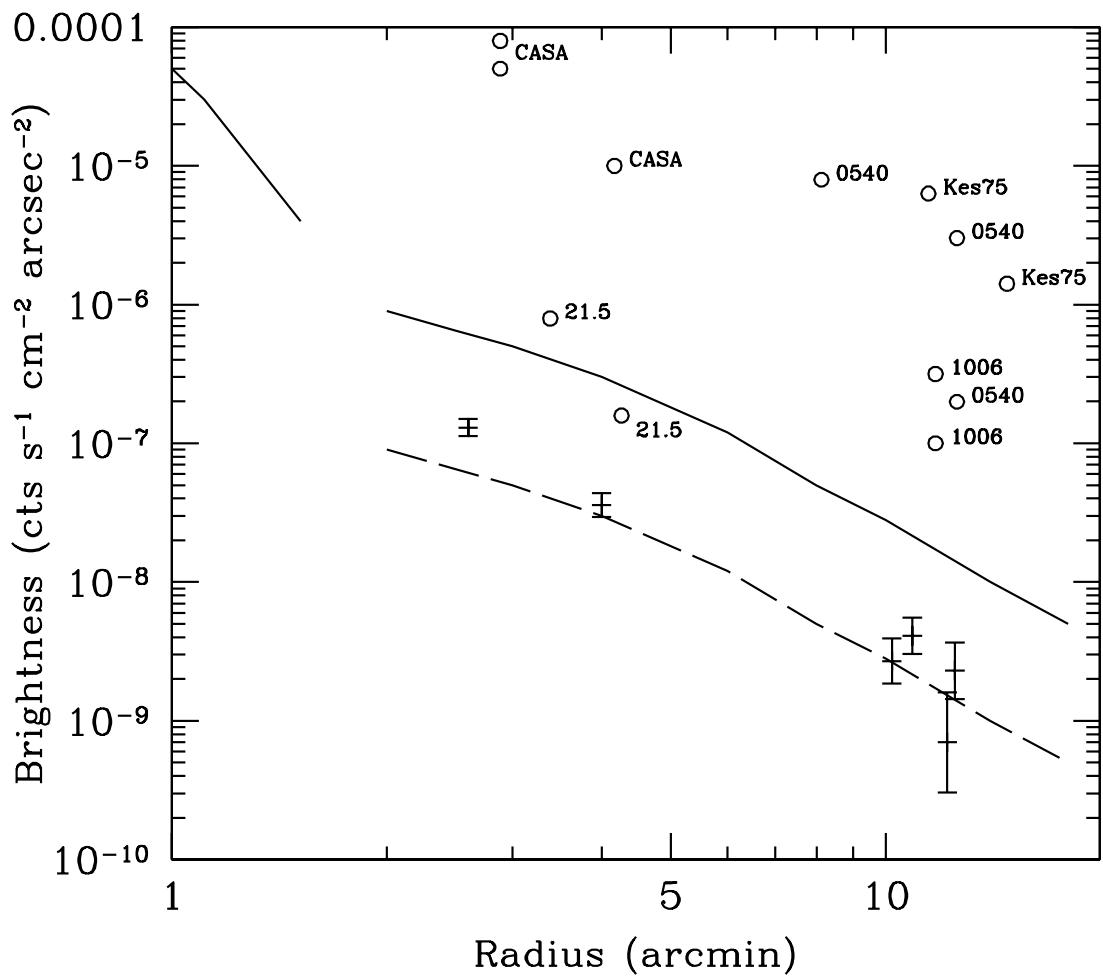}
\caption{Chandra-measured surface brightness around the Crab. The
dashed curve is 0.1 of the observed halo and is our threshold of
detection.  Crosses show some of the larger fluctuations in the
brightness pattern and illustrate that the dashed curve is a
reasonable detection threshold.  The cross at $2.5^{\prime}$ is a real 
feature, visible in Figure 4; others are statistical 
fluctuations (with number-of-counts
uncertainties).   Circles indicate approximate radii and
 brightness of other remnant shells if viewed from 2 kpc distance.}
\end{figure}

\begin{figure}
\center
\plotone{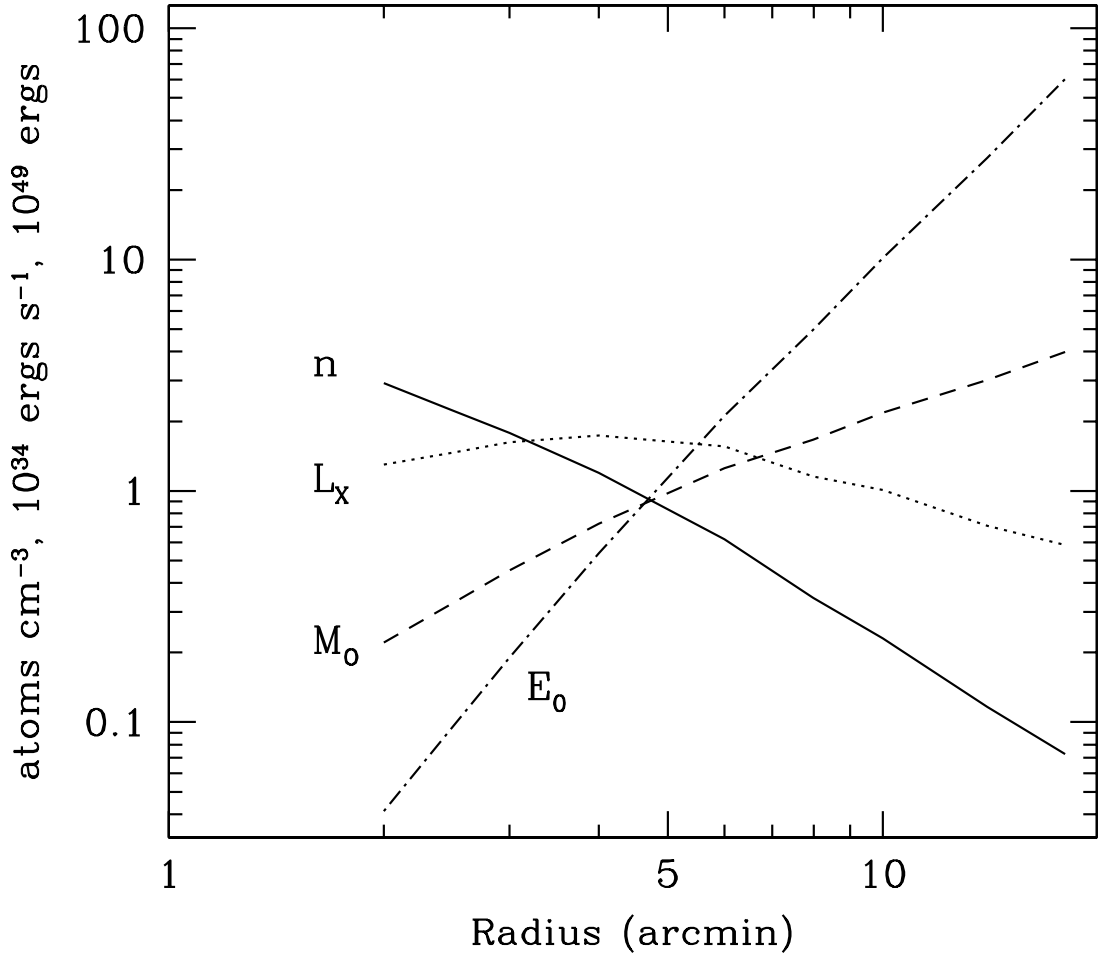}
\caption{Upper limits calculated for a uniform shell with
  brightness at the threshold of detection.}
\end{figure}

\begin{figure}[ht]
\plotone{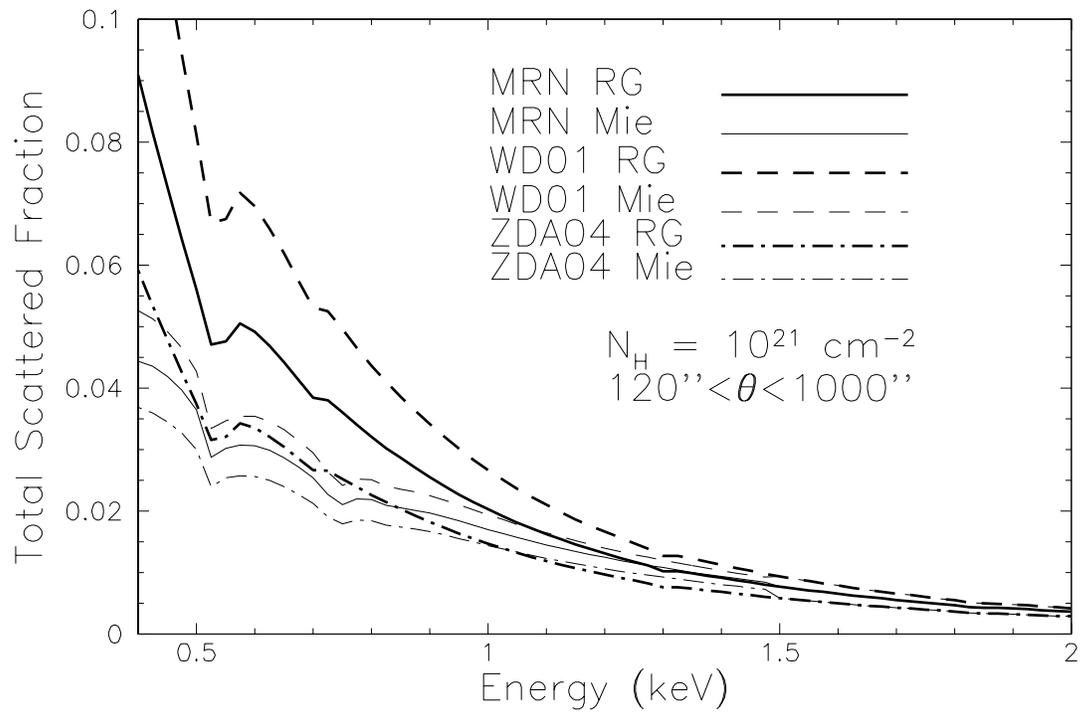}
\caption{The total scattering fraction as a function of energy between
$120-1000''$\ using three dust models and both the
Mie solution and the RG approximation.  Although there is a
significant difference between the models at low energies in the
RG approximation, the difference is much less when the Mie solution is
used.\label{fig:TotScatFrac}}
\end{figure}

\begin{figure}[ht]
\plotone{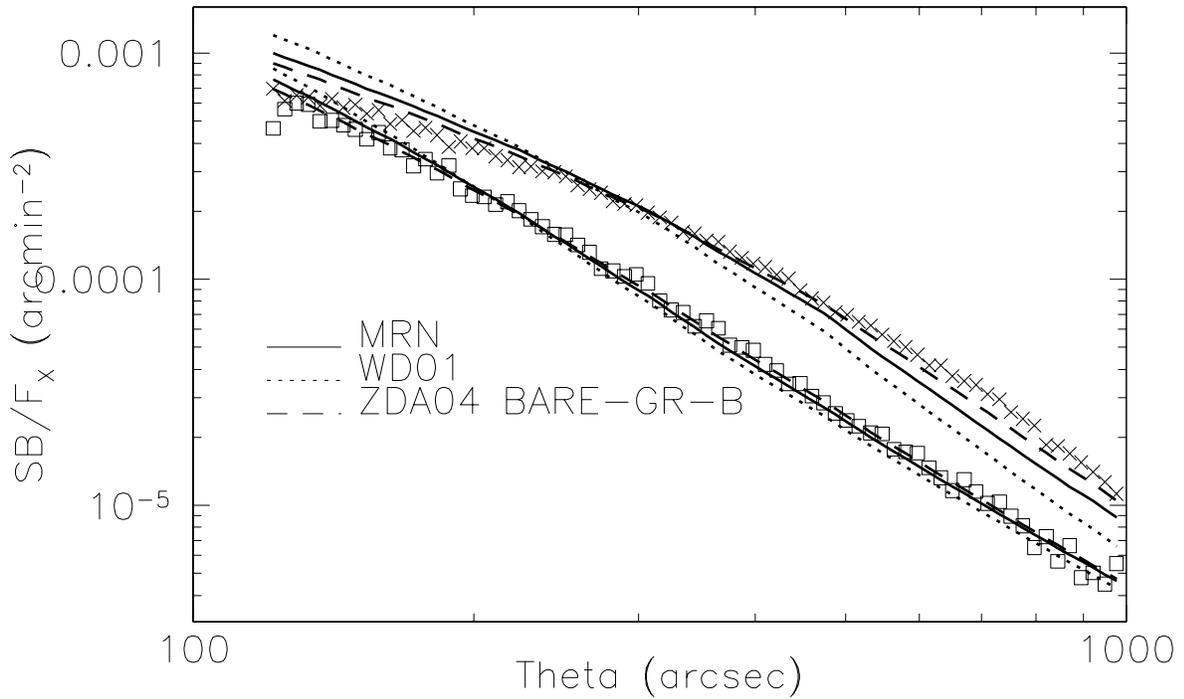}
\caption{Crab radial profiles at 1 ($\times$) and 2 ($\Box$) keV, 
fit with a smoothly-distributed 
MRN, WD01, and ZDA04 dust models.  The 1 keV fit used the Mie solution
and the the 2 keV fit the RG approximation.  At 1 keV, the ZDA04 
model is the best fit, although still poor; at 2 keV, the profile is 
dominated by mirror scattering with a weak dust halo in all three 
cases.\label{fig:SampleHalo}}
\end{figure}

\begin{figure}[ht]
\plotone{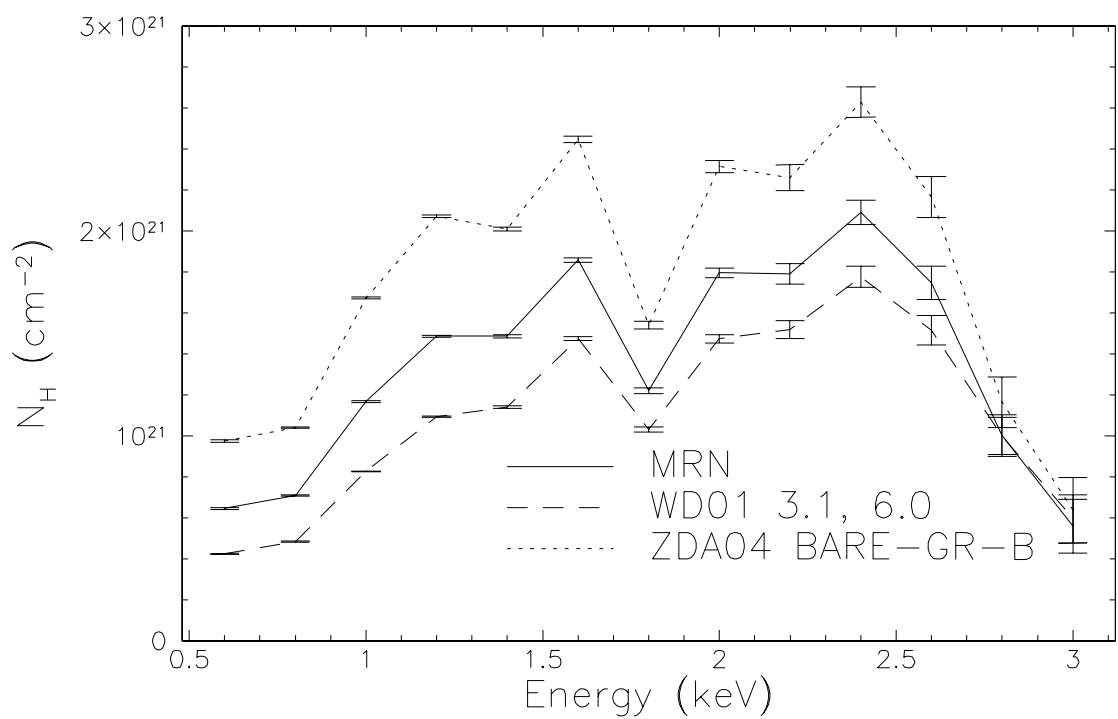}
\caption{The best-fit values of N$_{\rm H}$\ for the MRN, WD01 (using
  their $R_V = 3.1$, $Ab_C = 6.0$\ model), and ZDA04 (using their
  BARE-GR-B model) assuming a smooth spatial dust distribution.  Error
  bars show the statistical error only.  However, most of these fits
  have $2 < \chi^2 < 10$, implying that the errors are not purely
  statistical.  \label{fig:NHfit}}
\end{figure}

\begin{figure}[ht]
\plotone{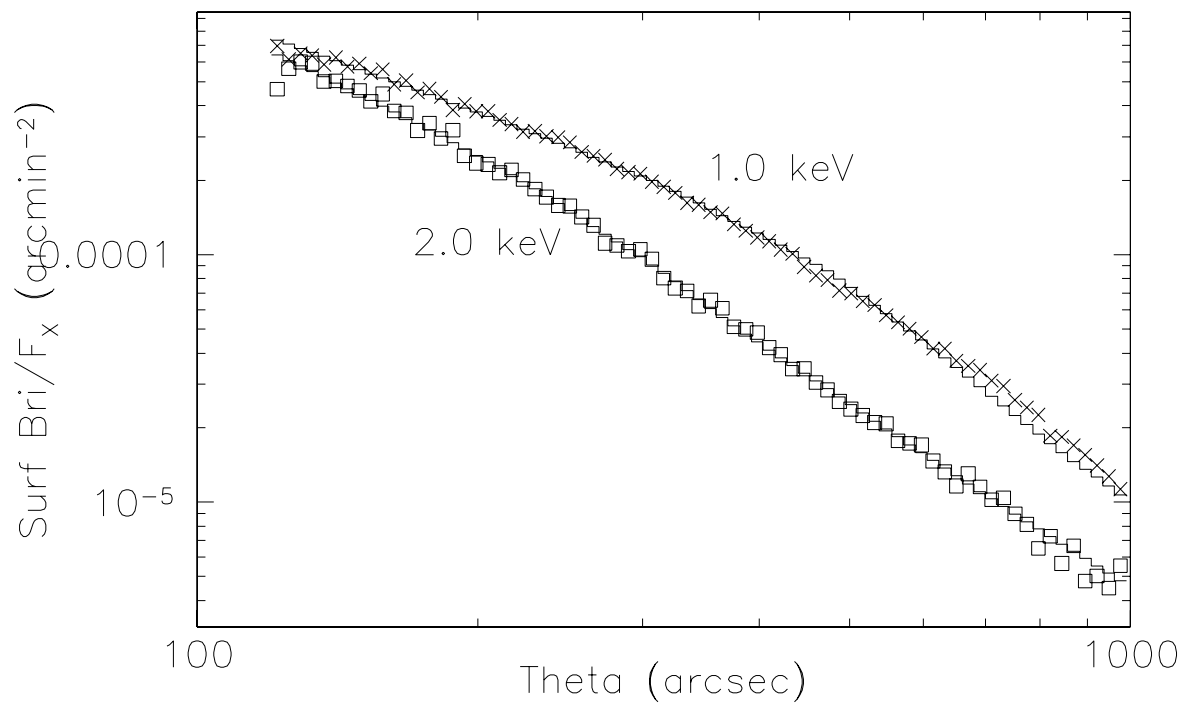}
\caption{Crab radial profiles at 1 and 2 keV, fit with a
  two-component MRN model with both smoothly-distributed plus a single
  dust cloud.  Data errors are approximately the size of the symbols.
  The best-fit column densities are $8\times10^{20}$\,cm$^{-2}$\ and
  $4\times10^{20}$\,cm$^{-2}$\ respectively for the 1 keV profile, and
  $9\times10^{20}$\,cm$^{-2}$, $5\times10^{20}$\,cm$^{-2}$\ at 2 keV.
  At both energies, the fit puts the dust cloud very near the
  Sun.}
\end{figure}

\begin{figure}[ht]
\plotone{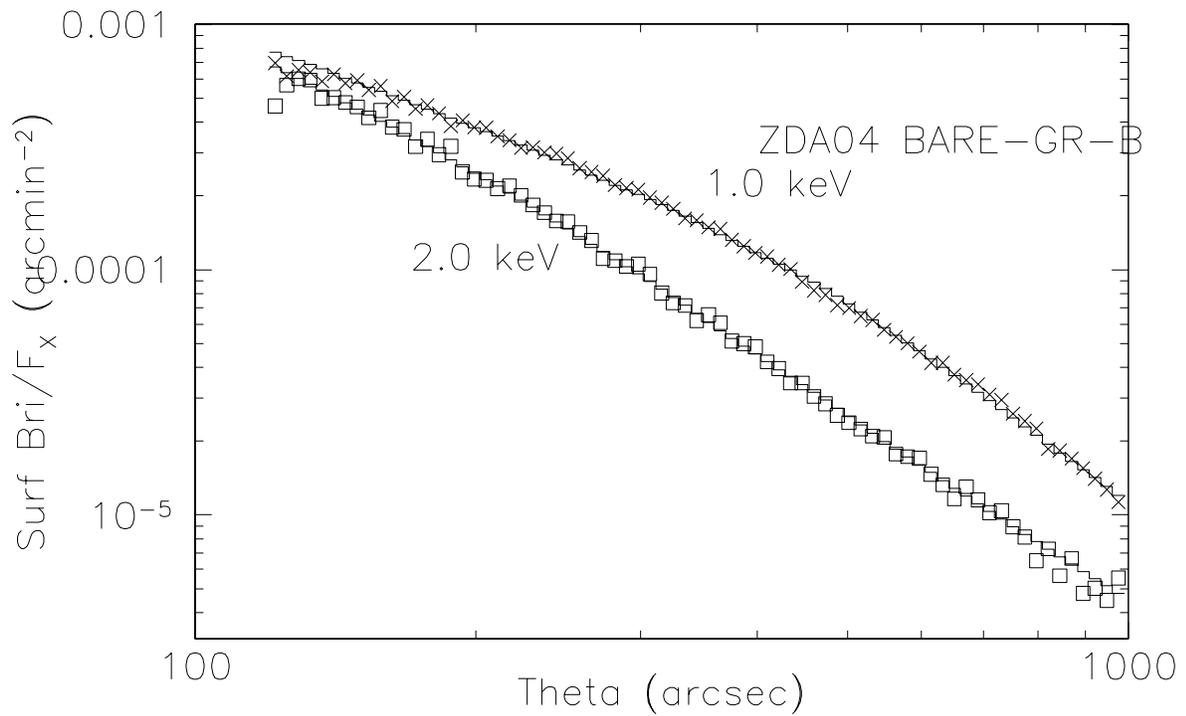}
\caption{Same as Figure 10, using a ZDA04 BARE-GR-B-type model.  In this case
  the best-fit column densities are slightly larger,
  $1.3\times10^{21}$\,cm$^{-2}$\ and $4\times10^{20}$\,cm$^{-2}$\ at 1
  keV, and $2\times10^{21}$, $2\times10^{20}$\,cm$^{-2}$\ at 2 keV.
  In this case, the best-fit cloud position is at 0.04 of the distance
  to the Crab, or $\sim 100$\,pc.}
\end{figure}

\begin{figure}
\center
\plotone{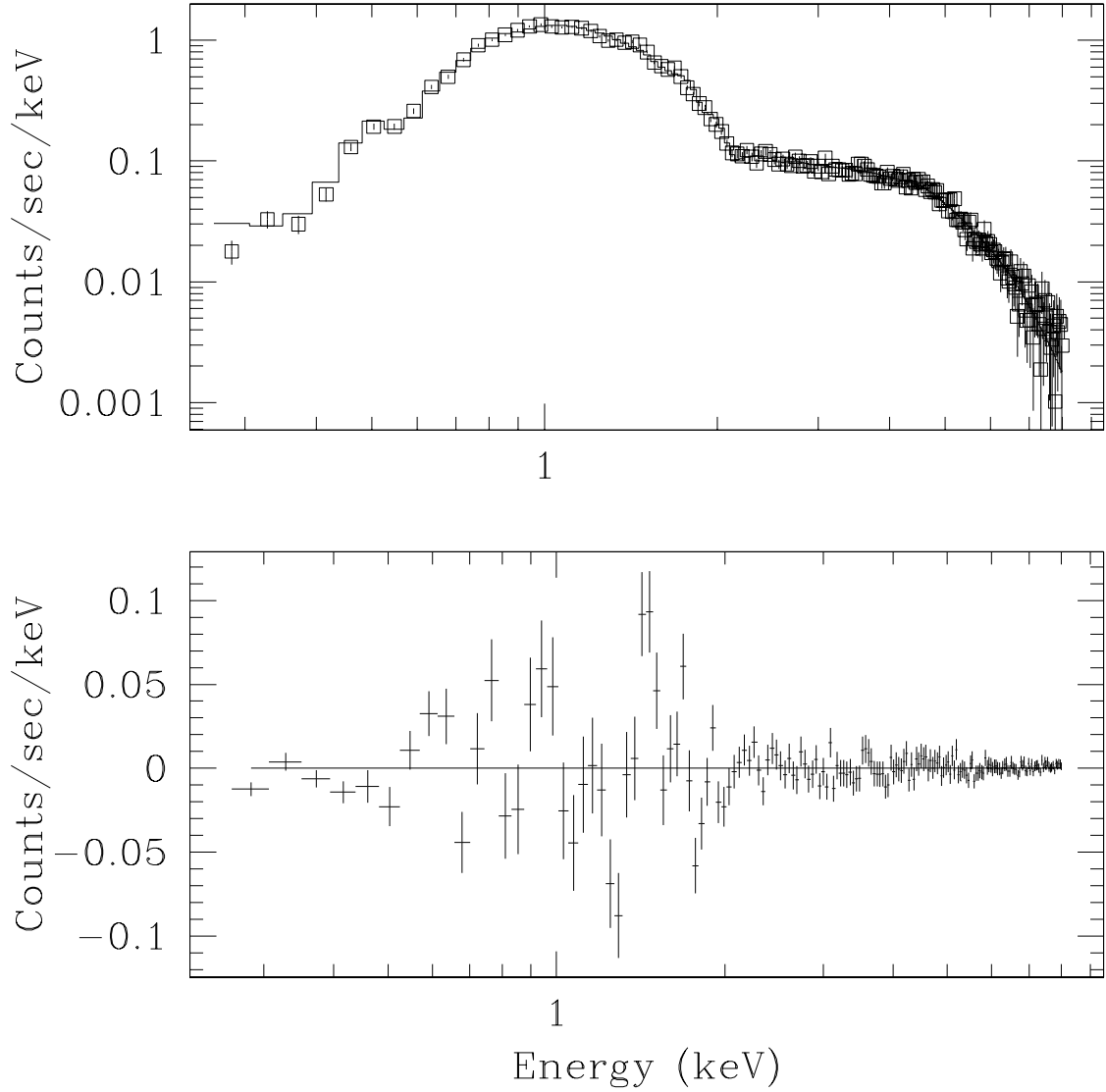}
\caption{X-ray spectrum of the halo 6$^{\prime}$ from the center
  of the nebula.  The fit is the sum of a broken power law and a thermal
  bremsstrahlung continuum}
\end{figure}

\begin{table}
\caption{Chandra observations}
\begin{tabular}{|l|l|l|l|} \hline\hline
observation number& date & live time& ACIS chips \\
\hline 
500174/1997 & 14 Mar 2001 & 3972 & S3 \\
500248/2798 & 14 Apr 2002 & 19981&I0,I1,I2,I3 \\
500432/4607 & 27 Jan 2004 & 37250 &S3 \\
500432/4607 & 27 Jan 2004 & 38090 &I2,I3,S1,S2 \\
\hline\hline 
\end{tabular} \\
\end{table}

\end{document}